\documentclass[aps,10pt,prl,a4paper,twocolumn,notitlepage,nofootinbib,floatfix]{revtex4-1}
\usepackage{amsmath}
\usepackage{graphicx}
\usepackage{color}
\usepackage[normalem]{ulem}

\begin{document} 

\title{General-relativistic rotation laws in rotating fluid bodies: constant linear velocity}

\author{Jerzy Knopik}
\author{Patryk Mach}
\author{Edward Malec}
\affiliation{Instytut Fizyki im.\ Mariana Smoluchowskiego, Uniwersytet Jagiello\'{n}ski, {\L}ojasiewicza 11, 30-348 Krak\'{o}w, Poland} 
 
\begin{abstract}
New rotation laws have been recently found for general-relativistic self-gravitating stationary fluids. It was not clear whether they apply to systems rotating with a constant linear velocity. In this paper we fill this gap. The answer is positive. That means, in particular, that these systems should  exhibit the recently discovered general-relativistic weak-field effects within rotating tori: the dynamic anti-dragging and the deviation from the Keplerian motion induced by the fluid selfgravity.
\end{abstract}

\maketitle

\section{Introduction}

Axially symmetric and stationary Newtonian hydrodynamic configurations are known for a long time to be  characterized by a rich variety of rotation curves. The  angular momentum per unit mass  $j$ can be any function of $r$, where $r$ is the distance from a (fixed) rotation axis.   In contrast to that, in general relativity had been known only two families of rotation laws---one with $j$ being a linear function of the angular velocity $\Omega$ \cite{Bardeen_1970, Butterworth_Ipser, komatsu} and a more recent nonlinear angular velocity proposal \cite{GYE}. Their Newtonian limits recover only a small fraction of the set of Newtonian rotation curves.  

Quite recently two of us have found  general-relativistic rotation curves $j = j(\Omega)$ \cite{MM} that in the nonrelativistic limit exactly coincide with all monomial rotation laws $\Omega_0=w/r^{2/(1-\delta)}$, with the  exception of the constant linear velocity case ($-\infty  \le \delta  \le 0$, $\delta \ne -1$, $w = \mathrm{const}$). We  obtained in particular the general-relativistic Keplerian rotation law that possesses the first post-Newtonian limit (1PN) and exactly encompasses the solution corresponding to the massless disk of dust in the Schwarzschild spacetime.

The rotation law proposed in \cite{MM} reads
\begin{equation}
\label{momentum}
j(\Omega) \equiv \frac{w^{1 - \delta}  \Omega^\delta}{1 - \frac{ 1 - 3 \delta}{(1 + \delta) c^2} w^{1 - \delta} \Omega^{1 + \delta} + \frac{4 c_0}{c^2}}.
\end{equation}

The main purpose of this paper is to show that the problematic case of constant linear velocity is also included in the proposed family of general-relativistic rotations.

\section{Hydrodynamical equations}

In this paper we apply the formulation of general-relativistic hydrodynamics elaborated by Komatsu et al. in \cite{komatsu}. Einstein equations read
\begin{equation}
R_{\mu \nu} -g_{\mu \nu }\frac{R
}{2} = 8 \pi \frac{G}{c^4} T_{\mu \nu},
\end{equation}
where $T_{\mu\nu}$ is the stress-energy tensor. We shall assume axial symmetry, stationary rotation and the angular velocity vector field $\vec{v}=\Omega\partial_{\phi}$. Then  the metric can be written as
\begin{eqnarray*}
ds^2 & = &  -  e^{\frac{2\nu }{c^2} }(d x^0)^2 + r^2 e^{\frac{2\beta }{c^2} } \left( d \phi - \frac{\omega}{c^3} \left( r, z \right) d x^0 \right)^2 \\
& & + e^{\frac{2 \alpha }{c^2}} \left( dr^2 + dz^2\right),
\end{eqnarray*}
where $r$, $z$, $\phi$ are the cylindrical coordinates. The stress-energy tensor of a relativistic perfect fluid reads
\begin{equation*}
T^{\alpha\beta} = \rho (c^2+h)u^\alpha u^\beta + p g^{\alpha\beta}.
\end{equation*}
Here $\rho$ is the baryonic rest-mass density, $h$ is the specific entalpy and $p$ is the pressure. The 4-velocity  {$u^\alpha  $}  is normalized, $g_{\alpha\beta}u^\alpha u^\beta=-1$ and $u^{\phi}/u^t=\Omega$.\par
We assume that the equation of state obeys a polytropic relation
\begin{equation*}
p(\rho,S)=K(S)\rho^{\gamma},
\end{equation*}
where $S$ is the specific entropy of the fluid and $\gamma$ is known as the adiabatic index. Hence $h(\rho, S)=K(S)\frac{\gamma}{\gamma-1}\rho^{\gamma-1}$. The  entropy is constant.

We introduce, following  \cite{komatsu},   
\begin{equation*}
V^2=r^2 \left( \Omega -\frac{\omega }{c^2}\right)^2 e^{2\left( \beta - \nu \right)/c^2},
\end{equation*}
where $V$ is the proper velocity with respect to the zero angular momentum observer.

Einstein equations implied by the above metric take the form of an overdetermined, but consistent set of equations imposed on the potentials $\alpha$, $\beta$, $\nu$, and $\omega$. These equations have been found by Komatsu et al.\ in \cite{komatsu}.

If we assume that the angular momentum per unit mass,
\begin{equation}
\label{momentum1}
j  =  u_\phi u^t= \frac{V^2}{\left( \Omega -\frac{\omega }{c^2}\right) \left( 1-\frac{V^2}{c^2}\right) },
\end{equation}
depends only on the angular velocity $\Omega$ [$j \equiv j(\Omega)$], then the Euler equations become solvable and they  reduce to  a single general-relativistic integro-algebraic Bernoulli equation
\begin{equation}
\label{Bernoulli}
\ln \left( 1+\frac{h}{c^2}\right) +\frac{\nu }{c^2} +\frac{1}{2}\ln \left( 1-\frac{V^2}{c^2}\right) +\frac{1}{c^2}\int d\Omega j(\Omega ) =C,
\end{equation}
where $C$ is an integration constant. The above equation carries all information that is present within the conservation equations  $\nabla_\mu T^{\mu \nu }=0$ and the baryonic mass conservation
$\nabla_\mu \left( \rho u^\mu \right) = 0$.

\section{Rotation law}

The general-relativistic rotation law employed in \cite{MM} equates the angular momentum per unit mass, given by (\ref{momentum1}), to a specific function $j(\Omega)$:
\begin{equation}
\label{eq: rotationlaw}
j(\Omega ) \equiv \frac{w^{1 - \delta}  \Omega^\delta}{1 - \frac{1 - 3\delta}{(1 + \delta) c^2}  w^{1 - \delta} \Omega^{1 + \delta} + \frac{4 c_0}{c^2}}.
  \end{equation}
In explicit terms one has  
\begin{equation}
\label{eq: rotation_law}
\frac{w^{1-\delta }  \Omega^\delta }{1-   \frac{ 1-3\delta   }{(1+\delta )   c^2 }  w^{1- \delta }\Omega^{1+\delta } +\frac{4 c_0}{   c^2}  } = \frac{V^2}{\left( \Omega -\frac{\omega }{c^2}\right) \left( 1-\frac{V^2}{c^2}\right) }.  
\end{equation}
From this equation one can recover rotation curves $\Omega(r,z)$.

With the rotation law (\ref{eq: rotationlaw}) the general-relativistic Bernoulli equation (\ref{Bernoulli}) acquires a simple algebraic form, assuming $\delta \neq -1$:
\begin{eqnarray}
\left( 1 + \frac{h}{c^2} \right)  e^{\nu/c^2} \sqrt{1-\frac{V^2}{c^2}} \times & & \nonumber  \\  
\left( 1 - \frac{1 - 3 \delta}{c^2 (1 + \delta)}  w^{1 - \delta} \Omega^{1 + \delta} + \frac{4 c_0}{c^2} \right)^{\frac{-1}{\left( 1 - 3\delta \right)}}  & = & C.
\label{algebraic_Bernoulli}
\end{eqnarray} 
  
Assume that there exists the Newtonian limit (the zeroth order of the post-Newtonian expansion---0PN) of the rotation law. This yields, in the 0PN order,
\begin{equation}
\label{zeroth_rotation_law}
\Omega_0 = \frac{w}{r^\frac{2}{1- \delta }}.
\end{equation}
Thus $w$ and $\delta $ can be obtained from the Newtonian limit.  Moreover, the  constant $w$ is any real number, while $\delta $ is nonpositive---due to the stability requirement \cite{Tassoul}---and satisfies the bounds $-\infty \le \delta \le  0$  and $\delta \neq -1$. These two constants can be given apriori within the given range of values.  Let us point out that the general-relativistic extension of this stability condition, formulated in \cite{KEH89}, is satisfied in our case for $\delta < 0$.

The two limiting cases $\delta  = 0$  and $\delta = - \infty$ correspond to the constant angular momentum per unit mass  ($\Omega_0 = w/r^2$) and the rigid rotation ($\Omega = w$), respectively. The Keplerian rotation is related to the choice of $\delta = -1/3$ and $w^2 = GM$, where $M$ is a mass.  

The particular form of the expression $(1 - 3\delta)/(1 + \delta)$ in the denominator of (\ref{eq: rotationlaw}) follows from the condition that a rotating infinitely thin disk made of weightless dust in a Schwarzschild space-time satisfies exactly the Bernoulli equation and the Keplerian rotation law \cite{MM}. 
 
It was proven in \cite{MM} that if  $c_0$ is the Newtonian hydrodynamic energy per unit mass, then the exact solution satisfies the first post-Newtonian (1PN) equations. We shall outline here the calculation.  

The 1PN approximation corresponds to the choice of metric exponents
$\alpha = \beta = -\nu = -U$, with $|U| \ll c^2$ \cite{BDS}. Define $\omega \equiv r^{-2} A_\phi$. The spatial part of the metric 
\begin{eqnarray}
ds^2 & = & - \left( 1 + \frac{2U}{c^2} + \frac{2U^2}{c^4} \right) (d x^0)^2 - 2  c^{-3} A_\phi d x^0 d \phi \nonumber \\
& & + \left( 1 - \frac{2U}{c^2} \right) \left( d r^2 + d z^2 + r^2 d \phi^2 \right). 
\label{metric1}
\end{eqnarray} 
is conformally flat.   
  
We split different quantities ($\rho$, $p$, $h$, $U$, and $v^i$) into their Newtonian (denoted by subscript `0' and 1PN (denoted by subscript `1') parts. Exempli gratia, for $\rho$, $\Omega$, $\Psi$, and $U$ this splitting reads
\begin{subequations}
\label{density_rotation}
\begin{align}
\rho & = \rho_0 + c^{-2} \rho_1, \\
\Omega & = \Omega_0  + c^{-2} v_1^\phi, \\
U & = U_0 + c^{-2} U_1.
\end{align}
\end{subequations}
Notice that, up to the 1PN order,   
\begin{equation}
\label{enthalpy}
\frac{1}{\rho} \partial_i p = \partial_i h_0 + c^{-2} \partial_ih_1
 {+ \mathcal{O}(c^{-4})},
\end{equation}
where the {1PN} correction $h_1$ to the specific enthalpy can be written as $h_1 = \frac{dh_0}{d\rho_0} \rho_1$. For the polytropic equation of state this gives $h_1 = \left( \gamma -1 \right) h_0 \rho_1 / \rho_0$.
 
Making use of the introduced above splitting of quantities into Newtonian 0PN and 1PN parts
one can extract from Eq.~(\ref{algebraic_Bernoulli}) the 0PN- and 1PN-level Bernoulli equations.

At the Newtonian level the gravitational potential is given by the Poisson equation  
\begin{equation}
\label{DeltaU0}
\Delta U_0 = 4 \pi G \rho_0,
\end{equation}
while the Bernoulli equation reads
\begin{equation}
\label{0Bernoulli}
h_0 + U_0 - \frac{\delta - 1}{2(1+ \delta )} \Omega^2_0 r^2 = c_0,
\end{equation}
where $c_0$ is a constant that can be interpreted as the energy per unit mass. Here $\Delta$ is the flat Laplacian with respect to the cylindrical coordinates $r$, $z$, and $\phi$.
 
One can obtain the first correction $v_1^{\phi}$ to the angular velocity $\Omega$ by expanding  the rotation law (\ref{eq: rotation_law}) in powers of $c$ up to terms $\mathcal(c^{-2})$
\begin{equation}
v_{1}^{\phi} = - \frac{2}{1 - \delta}\Omega^3_0 r^2 + \frac{A_{\phi}}{r^2(1 - \delta)} - \frac{4 \Omega_0 h_0}{1 - \delta}.
\label{eq: velcorrection}
\end{equation}  
Using the fact that in the Newtonian gauge imposed on the line element (\ref{metric1}) the distance to the rotation axis measures $\tilde{r} = r \left(1 - U_0/c^2\right) + \mathcal{O}(c^{-4})$, we can write down the full expression for the angular velocity, up to terms $\mathcal{O}(c^{-4})$ \cite{MM}
\begin{eqnarray}
\label{correction}
\Omega & = & \Omega_0 + \frac{v_1^{\phi}}{c^2} = \nonumber \\
& & \frac{w}{\tilde{r}^{2/(1 - \delta)}} - \frac{2}{c^2(1 - \delta)} \Omega_0 \left( U_0 + \Omega_0^2 \tilde r^2 \right) + \nonumber \\
& & + \frac{A_{\phi}}{\tilde r^2c^2 (1 - \delta)} - \frac{4}{c^2(1 - \delta)}\Omega_0 h_0.
\end{eqnarray}

\section{Constant linear velocity}

In order to construct the limit $\delta\rightarrow-1$, we need to inspect the denominator of (\ref{eq: rotationlaw}). One can easily check, using (\ref{0Bernoulli}), that in this limit both $c_0$ and $\frac{1 - 3\delta}{c^2 (1 + \delta)} w^{1 - \delta}\Omega^{1 + \delta}$ become singular. After addition and subtraction of a term $\frac{2(\delta - 1)}{c^2(1 + \delta)}w^{1 - \delta}$ in the denominator of (\ref{eq: rotationlaw}) we obtain
\begin{align}
j(\Omega ) & \equiv \frac{w^{1 - \delta } \Omega^\delta}{1 - \frac{1 - 3 \delta}{(1+\delta )   c^2 }  w^{1- \delta }\Omega^{1+\delta } -\frac{2(\delta -1)}{c^2(1+\delta)}w^{1-\delta}+\frac{4 \hat{c}_0}{   c^2}  } \nonumber \\
& = \frac{w^{1-\delta }  \Omega^\delta }{1+\frac{1}{c^2}w^{1-\delta}\Omega^{1+\delta}-\frac{2(1-\delta)}{c^2}w^{1 - \delta}\left(\frac{\Omega^{1+\delta}-1}{1+\delta}\right)+\frac{4 \hat{c}_0}{c^2}},
\label{eq: reg_rotationlaw}
\end{align}
where $\hat{c}_0$ is the regularised energy per unit mass 
\begin{align}
\hat c_0 & = h_0 + U_0 - \frac{\delta - 1}{2(1 + \delta)} \Omega^{1 + \delta}w^{1 - \delta} + \frac{\delta - 1}{2(1 + \delta)} w^{1 - \delta} \nonumber \\
& = h_0 + U_0 -\frac{\delta - 1}{2} w^{1 - \delta} \left(\frac{\Omega^{1 + \delta} - 1}{1 + \delta}\right).
\label{eq: reg_const}
\end{align}
It is worth noting that such defined $\hat{c}_0$ and $j(\Omega)$ are not singular in $\delta = -1$. To prove this fact, one needs only to use the identity
\begin{equation*}
\lim_{\alpha\rightarrow 0}\frac{x^{\alpha} - 1}{\alpha} = \ln x.
\end{equation*}
Indeed, taking the limit $\delta \rightarrow - 1$ in equations (\ref{eq: reg_rotationlaw}) and (\ref{eq: reg_const}) results in
\begin{equation*}
j(\Omega ) = \frac{w^{2} \Omega^{-1}}{1 + \frac{1}{c^2}w^{2} - \frac{4}{c^2} w^{2} \ln \Omega + \frac{4 \hat{c}_0}{c^2}},
\end{equation*}
while the regularised Bernoulli equation reads now
\begin{equation}
\hat c_0 = h_0 + U_0 + w^2 \ln \Omega.
\label{reg0Bernoulli}
\end{equation}  
 
As in the case $\delta \ne -1$, the leading correction $v^\phi_1$ to the angular velocity $\Omega_0$ is obtained from the perturbation expansion of the rotation law
\begin{equation}
\label{reg_rotation_law}
\frac{w^{2} \Omega^{-1}}{1 + \frac{w^2}{c^2} - \frac{4 w^2 \ln \Omega}{c^2} + \frac{4 \hat c_0}{c^2}} = \frac{V^2}{\left( \Omega -\frac{\omega}{c^2}\right) \left( 1 - \frac{V^2}{c^2}\right)},  
\end{equation}
up to terms of the order $c^{-2}$. One arrives at
\begin{equation}
\label{reg_constraint_solution_th}
v^\phi_1 = - \Omega_0^3 r^2 + \frac{A_\phi}{2r^2} - 2\Omega_0 h_0,
\end{equation}
where we applied Eqs.~(\ref{zeroth_rotation_law}) and (\ref{reg0Bernoulli}). Note that the form of the corection is not affected by the transition to the limit and agrees with (\ref{eq: velcorrection}).
  
After these consideration we are able to interpret the meaning of various contributions to the angular velocity $\Omega$ given by formula  (\ref{correction}). The first term is simply the Newtonian rotation law rewritten as a function of the geometric distance, as given at the 1PN level of approximation, from the rotation axis. The second term in (\ref{correction}) is sensitive  both to the contribution of the disk self-gravity at the plane $z=0$ and the deviation  from the strictly Keplerian motion. It vanishes for test fluids. The third  term is responsible for the geometric frame dragging. The last term represents the recently discovered dynamic anti-dragging effect; it agrees (for the monomial angular velocities $\Omega_0 =r^{-2/(1 - \delta )}w$)---with the result obtained earlier in \cite{JMMP}.  
 
We shall consider the remaining first order perturbations terms.
 
The vectorial component $A_\phi$ satisfies the following equation
\begin{equation}
\label{Afi}
\Delta A_\phi -2 \frac{\partial_r A_\phi }{r} = -16 \pi G r^2 \rho_0 \Omega_0.
\end{equation}

The 1PN Bernoulli equation has the form  
\begin{eqnarray}
c_1 & = & -h_1 - U_1 -\Omega_0 A_\phi + 2 r^2 (\Omega_0)^2 h_0 - \frac{3}{2} h^2_0 \nonumber \\
& & - 4 h_0 U_0 - 2 U_0^2 - \frac{\delta + 3}{4\left( 1 + \delta \right)} r^4\Omega^4_0 \nonumber \\
& = & -h_1 - U_1 -\Omega_0 A_\phi + 2 r^2 (\Omega_0)^2 h_0 - \frac{3}{2} h^2_0 \nonumber \\
& & - 4 h_0 U_0 - 2 U_0^2 - \frac{(\delta + 3) w^4}{4\left( 1 + \delta \right)},
\label{Psi}
\end{eqnarray}
where $c_1$ is a constant.
 
The 1PN potential correction $U_1$ can be obtained from   
\begin{equation}
\label{DeltaU1}
\Delta U_1 = 4 \pi G \left( \rho_1 + 2 p_0 + \rho_0(h_0 - 2 U_0 + 2 r^2(\Omega_0)^2) \right).
\end{equation}

It is clear that only the correction to the proper energy per unit mass $c_1$ would cause trouble in the limit $\delta \rightarrow -1$, but on the other hand it  does not influence the 1PN correction to the angular velocity. One would have to regularise $c_1$ in the next orders of the perturbation calculation.

\section{Summary}

We found in \cite{MM} that the consistency of the formalism  and the uniqueness of those solutions that possess the 1st post-Newtonian limit, lead to well defined  numerical values of the coefficients in the  rotation law (\ref{momentum}). It is interesting  that the definition (\ref{momentum}) is rigid---there is no any parametric freedom left for those solutions that have the 1st post-Newtonian limit.  We proved in this paper that the case of constant linear velocity, that  corresponds to the seemingly singular point $\delta = - 1$, can be deduced from (\ref{momentum}). That extends the validity of the general relativistic laws constructed in \cite{MM}. 
 
\section{Ackonwledgements}

PM acknowledges the support of the  Polish Ministry of Science and Higher Education grant IP2012~000172 (Iuventus Plus).

\end{document}